\documentclass[aps,prl,letterpaper,twocolumn,nofootinbib,floatfix,showpacs,groupedaddress,superscriptaddress]{revtex4}
\usepackage{graphics,graphicx,bbm,bm,times,color,amsfonts,amssymb,amsmath,epsfig}

\begin{document}
\title{Correlation effects in a discrete quantum random walk}
\author{J. B. Stang}
\affiliation{Institute for Quantum Information Science, University
of Calgary, Alberta T2N 1N4, Canada}
\author{A. T. Rezakhani}
\affiliation{Institute for Quantum Information Science, University of Calgary, Alberta T2N 1N4,
Canada}
\affiliation{Center for Quantum Information Science and Technology, Departments
of Physics and Chemistry, University of Southern California, Los Angeles, CA 90089, USA}
\author{B. C. Sanders}
\affiliation{Institute for Quantum Information Science, University
of Calgary, Alberta T2N 1N4, Canada}

\begin{abstract}
We introduce history-dependent discrete-time quantum random
walk models by adding uncorrelated memory terms and also by modifying Hamiltonian of the walker to include couplings with
memory-keeping agents. We next numerically study the correlation effects in these models.
We also propose a correlation exponent as a relevant and promising tool for investigation
of correlation or memory (hence non-Markovian) effects. Our analysis can easily be applied to more realistic models in
which different regimes may emerge because of competition between
different underlying physical mechanisms.
\end{abstract}
\pacs{03.67.-a, 05.40.Fb, 03.65.Yz}
\maketitle

\textit{Introduction.}--- ``Random walks" (RWs) is an important and
prevalent concept in various branches of science \cite{books}, in
that many phenomena can be modeled by using the associated notions or tools. A classical
random walk (CRW) is the dynamics of a classical (i.e., non-quantum)
object --- which will be called ``particle" or ``walker" hereon ---
within a fully or partially stochastic environment and/or under some
stochastic forces. A famous example of such an evolution is encountered in
non-equilibrium statistical mechanics of gas particles inside a
cylinder. Configuration or state of the walker in a CRW can be
described by a local (classically measurable) quantity, such as its
``position'' (not necessarily real-space position) at each step (or
``time"), $x(t)$. A general, and therefore, context-free modeling of
a CRW is provided when one can introduce stochasticity/randomness
via inclusion of a random object such as a ``coin". A characteristic
of a RW is its variance or dispersion, $\sigma^2=\langle
x(t)^2\rangle-\langle x(t)\rangle^2$, where $\langle . \rangle$
indicates an ensemble average. For a CRW, this quantity for long
times shows a linear behavior with the total walk time:
$\sigma^2\sim T$, which is a characteristic of a \textit{diffusive}
motion. More generalizations of CRW can be found in literature, for example see
Refs.~\cite{hara79,stanley83,hilfer91,ordemann00,tan02,hod05} and
the references therein.

Recently, there has been a great interest in the dynamics of a
\textit{quantum} random walk (QRW), a quantum object hopping (discretely or
continuously) on a graph --- e.g., a line --- based on an
intrinsically quantum mechanical decision-making in each step, e.g.,
by a quantum coin in a discrete QRW
\cite{aharonov93,meyer,watrous,farhi98,nayak00,childs03,childs04,mackay02,kempe03,strauch06}.
Moreover, it has been shown that the language of a (continuous) QRW,
assisted with suitable Hamiltonian maps, can provide a universal
framework for the studies on general qubit systems
\cite{hines-stamp}. One should note that in this type of RWs an
external object like a quantum coin is not necessary. In a QRW, the
combined dynamics of the coin and the walker is governed by a
unitary operation $U_{\text{CW}}$ (acting on
$\mathcal{H}_{\text{C}}\otimes\mathcal{H}_{\text{W}}$), which
introduces quantum effects such as coherence and entanglement and
results in interference between classical paths. This quantum nature
is responsible for the features radically different than those of a
CRW, such as: a different, spread non-Gaussian probability
distribution $P(x,t)$ \cite{nayak00,kempe03}, a quadratically faster
spreading $\sigma^2\sim T^2$
\cite{nayak00,kempe03,romanelli03,romanelli04} (the
\textit{ballistic} motion), the exponentially faster propagation
between particular nodes of a specific graph
\cite{farhi98,childs03}.

Of our special interest in this paper is to numerically investigate how
``memory-effects" or ``correlations" show up and play a role in the
general behavior of a (discrete, coined) QRW. In a RW, the dynamics
in every time is generally dictated by the history of the previous
step(s) and the coin-flip(s). When the dynamics is Markovian, in
principle there is no history in the system, and, the walker's
immediate future is decided only based on its present and an
immediate coin-flip \cite{books}. In other words, in a Markovian
CRW, by definition, the walker does not keep any memory of its state
in previous times. From a physical point of view, it seems that when
the walker is interacting with a slowly-responding environment ---
slow relative to the characteristic time of the walk --- it is
unlikely that the environment can feed some of the acquired (or
leaked) information back to the walker, and therefore, affect its
future moves. In this case, the leading effect would be a loss of
memory, and accordingly, emergence of a regime in which the Markovian
assumption is a valid approximation. As a result, it is expected
that in this regime, there would be a negligible correlation between
the configurations in distant times. In open quantum systems the
analysis of non-Markovian effects is more involved than in the classical
case and adding a memory kernel to equations has many complex
aspects \cite{breuer06}. Besides, unlike the classical case, in open
quantum systems one cannot use the standard approaches like the
Chapman-Kolmogorov equation to test the Markovian property
\cite{friedrich}, because the related joint probabilities might be
not well-defined from quantum mechanical perspective. Recently,
however, some preliminary progress was reported regarding how to decide
whether a quantum channel is Markovian or not \cite{wolf}.

In a discrete QRW, the stepwise coin-walker dynamics, i.e.,
$|\Psi(t+1)\rangle_{\text{CW}}=U_{\text{CW}}|\Psi(t)\rangle_{\text{CW}}$
for $t\leqslant T$, may imply a Markovian characteristic for the
walker's dynamics as well. Although for the coin-walker system the
dynamics is indeed Markovian (or memoryless), this is not generally
the case for the walker's state alone \cite{bracken04}. It has been
shown that due to the quantum entanglement between the coin's
and the walker's states, there can exist a ``pseudo" memory (hence
non-Markovian effect) in the RW after tracing out over the coin at
the last step (or after a few steps). Precisely speaking, in the case of
a standard discrete-time QRW with a localized initial state, the
quantum probability distribution $P_{\text{quant.}}(x,t)$ is related
to all classical probabilities
$\{P_{\text{class.}}(x,t'):~t'\leqslant t\}$ (with the same initial
conditions). Tracing over the coin immediately after each step, and
averaging over all possible measurement results, generates a CRW
\cite{brun03a}. This is typical of systems under decoherence or
interaction with an external environment. When the environment observes
or measures the system, some coherence, hence dynamical information,
would become inaccessible and the system tends to lose its quantum
correlations. This has been anticipated as a usual route to manifest
classical-like behavior in quantum systems. In the case of a QRW,
there are numerous studies specializing on how different sources of
decoherence can affect a QRW and induce a transition to a CRW. E.g.,
tracing over the coin after each step or replacing the used coin
with a new one at each step (multiple coins) and tracing over all of
them after a while \cite{brun03a,brun03b,flitney04}, random phase
noise on the coin state \cite{mackay02,kosik06}, unitary stochastic
noise \cite{shapira03}, periodic coin and/or walker measurements or
randomly broken links on the graph \cite{romanelli05} --- for a
general review see Ref.~\cite{kendon-review}. In these studies, the
long-time behavior of variance $\sigma^2$, for its different
scalings for the standard CRW and QRW (as discussed earlier), has
been adopted as an indicator to distinguish ``classical" and
``quantum" regimes of a RW \cite{brun03a}. This approach, although
very fruitful, is not necessarily conclusive in that there are
\textit{quantum} diffusion models featuring sub-ballistic,  the
so-called ``anomalous" diffusion, or other types of behaviors
\cite{zhong95,zhong-prl-01,mukho07,stamp}. This implies that a
deeper characterization of different regimes in quantum systems by
other stronger tools is necessary. Moreover, there
is still no clear understanding about possible roles memory,
correlations, or related environmental effects might play in
appearance of different regimes in an open system QRW or transitions
between such regimes. A study in this line, therefore, might shed some light
and bridge between seemingly different underlying notions and
physical behaviors. Here we report a numerical preliminary step
that may fill some blanks.

In the following, we introduce a few simple QRW models in which a
memory/history-keeping feature is included. First we add a
non-Markovian property to a QRW as an uncorrelated mixing of the
states at different instants. We show how variance for these models
behave vs time, signaling the inadequacy of this quantity for
distinguishing different regimes. Next, we consider a more physically
motivated model in which, in addition to the coin and the walker, a
simple harmonic oscillator has been coupled to play as a
history-keeping agent. We define the concepts of correlation and
correlation exponent as useful tools for evaluating the effect of
memory. The correlation exponent for our model is calculated
numerically and contrasted with the exponents of a memory-dependent
CRW model \cite{tan02}. This analysis implies that adding memory may
induce anti-correlation similar to what is seen in a self-avoiding CRW.

\textit{Uncorrelated history-dependence.}--- We start by a brief
review of the standard (memoryless, discrete) QRW \cite{kempe03}.
This model consists of a finite one-dimensional integer lattice
forming the walker's space,
$\mathcal{H}_{\text{W}}=\text{span}\lbrace|x\rangle\rbrace_{x=-L}^L$,
and a chirality (or spin) degree of freedom,
$\mathcal{H}_{\text{C}}=\text{span}\lbrace|\pm\rangle\rbrace$,
constituting the coin space. The dynamics of
$|\Psi(t)\rangle_{\text{CW}}$ is induced by the unitary operator
$U_{\text{CW}}(p)=(P_{+}\otimes S+P_{-}\otimes
S^{\dagger})(u_{\text{C}}(p)\otimes\openone_{\text{W}})$, where
$P_{\pm}$ are projection operators onto $\mathcal{H}_{\text{C}}$,
$S=\sum_{x=-L}^{L-1}|x+1\rangle_{\text{W}}\langle x|$ is the shift
operator on the graph, and $u_{\text{C}}(p)=\left(
\begin{smallmatrix}\sqrt{p} & \sqrt{1-p}\\ \sqrt{1-p} & -\sqrt{p}
\end{smallmatrix}\right)$, $0\leqslant p\leqslant 1$, is a unitary quantum coin tossing operator.

Intuitively, one might imagine that a memory-dependent model of a
QRW can be built by considering that the state
$|\Psi(t+M)\rangle_{\text{CW}}$ is obtained from some operation on a
linear combination of the states in the $M$ previous instants:
$\lbrace|\Psi(t+m)\rangle_{\text{CW}}\rbrace_{m=0}^{M-1}$. This
simple approach is unfortunately doomed to be non-linear hence
unphysical, though. There are various ways to avoid non-linearity. Here
we consider two simple (though not necessarily
physically motivated) models in which the states at different
instants are mixed in an uncorrelated and random manner. A rather
similar approach has already been used to numerically investigate
decoherence effects on a QRW \cite{maloyer}.

\begin{figure}[tp]
 \includegraphics[width=8.5cm]{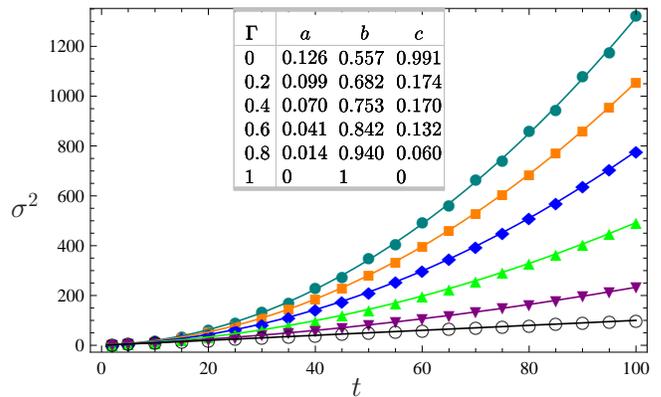}
\caption{(color online). Variance vs time for the model in
Eq.~(\ref{M0}), with $M=2$, $\Gamma_1=\Gamma$, and
$\Gamma_2=1-\Gamma$, $T=100$ and $p=1/2$. The inset shows the values
of $\Gamma$ and the parameters in the fitting $\sigma^2\sim
at^2+bt+c$.} \label{M0plot}
\end{figure}
\begin{figure}[tp]
\includegraphics[width=8.5cm]{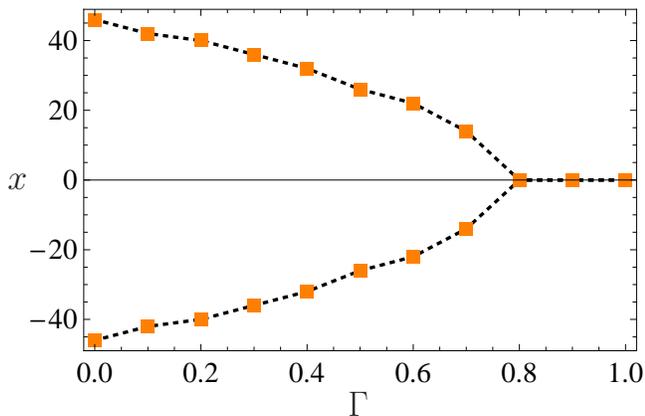}
\caption{(color online). Positions of the peaks of the probability
distribution at $t=100$ for the evolution described in
Eq.~\eqref{M0}. This model exhibits bimodal (unimodal) behavior
for $0\leqslant\Gamma<0.8$ ($\Gamma>0.8$).} \label{bifurcation}
\end{figure}

The first model is based on the density matrices in the $M$ previous steps \begin{eqnarray}
&\rho_{\text{W}}(t+M)=\sum_{\sigma=\pm,k=1}^{M}\Gamma_k A_{\sigma}^{(k)}\rho_{\text{W}}(t+M-k) A_{\sigma}^{(k)\dag}.
\label{M0}
\end{eqnarray}
Here, $0\leqslant\Gamma_k\leqslant 1$ and $\sum_k\Gamma_k=1$ and $A_{\sigma}^{(k)}$ is the Kraus operator given by
$A_{\sigma}^{(k)}= \langle\sigma|U_{\text{CW}}^k|C\rangle$, where
$|C\rangle=C_{+}|+\rangle + C_{-}|-\rangle$ is the initial state of
the coin. Non-Markovian characteristic of this model is apparent for
the history-dependence of the walker's state on its $M$ previous
instances. Another feature inferred from Eq.~(\ref{M0}) is that in
every step we discard the coin, tracing out over it to obtain
$\rho_\text{W}$, and use a fresh coin prepared as $|C\rangle$ for
the next step (multiple coins). Indeed, including multiple coins has
already been identified as a way to include history in a QRW
\cite{flitney04}. For a balanced initial coin
($C_+=-iC_-=1/\sqrt{2}$) and a balanced coin-flip ($p=1/2$), if
$\Gamma_1=1$ we obtain a behavior similar to an unbiased memoryless
CRW in the sense of variance (see Fig.~\ref{M0plot}). Degree of
history-dependence of the model is adjusted by $\Gamma_k$s. For
numerical simulations, in addition to the aforementioned conditions,
we have taken $M=2$ and $T=100$, with the walker initially localized
at the origin. Figure~\ref{bifurcation} depicts the peaks of the
probability distribution $P(x,100)$ for varying $\Gamma$. Notice
that for a range of values of $\Gamma$, the probability distribution
exhibits a bimodal behavior, similar to a memoryless QRW. As
$\Gamma$ increases, the the peaks approach each other and eventually
merge into one for $\Gamma\approx 0.8$. As indicated in the inset of
Fig.~\ref{M0plot}, for this walk a $\sigma^2\sim
a(\Gamma)t^2+b(\Gamma)t+c(\Gamma)$ fitting can be found for
different values of $\Gamma$ --- with $\sigma^2\sim t$ for
$\Gamma=1$. In this respect this model shows characteristics of both
QRW and CRW for different ranges of $\Gamma$.

\begin{figure*}[tp]
\includegraphics[scale=.5]{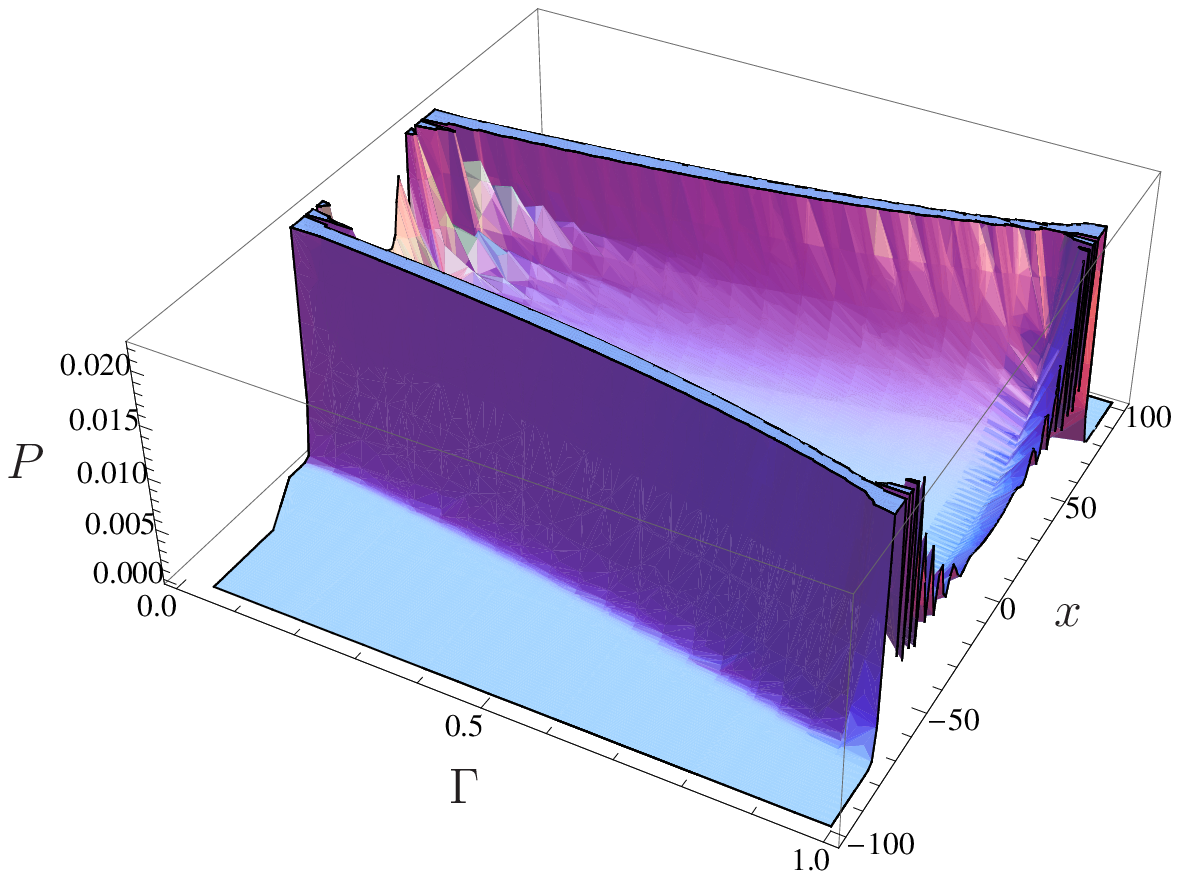}\includegraphics[scale=.5]{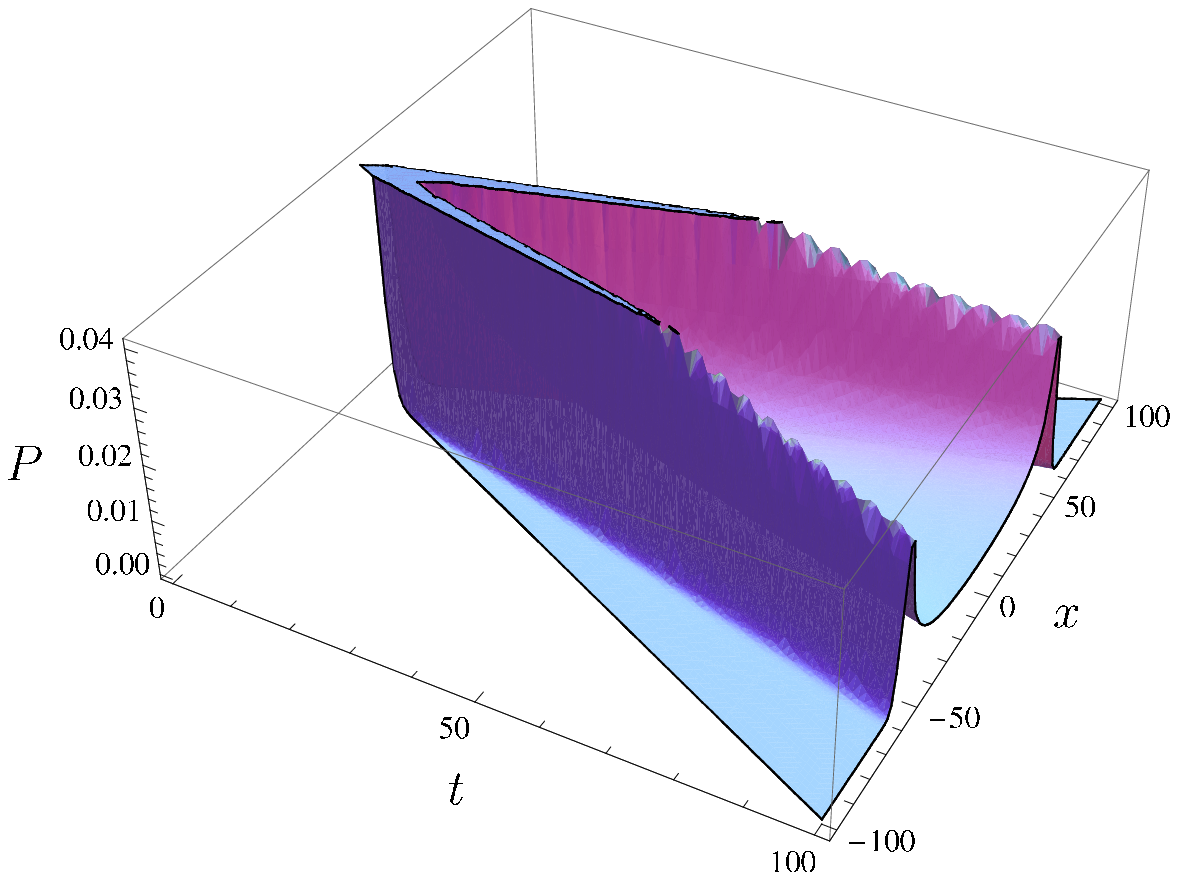}
\caption{(color online). (Left) Probability distribution $P(x,100)$
vs $\Gamma$ for Eq.~(\ref{M2}) and (right) $P(x,t)$ for
$\Gamma=2/3$. In the both  plots we have taken $p=1/2$. }
\label{m2-combined}
\end{figure*}

The second model is an uncorrelated mixing of some unitary dynamics
which overall constitutes a dependence on information from the $M$
previous steps. Using $M$ different quantum coin-walk operators,
$U_\text{CW}(p_k)\equiv U_\text{CW}^{(k)}$ (for notational
convenience), generated by $\lbrace p_k\rbrace_{k=1}^M$, the
evolution is described as the following:
\begin{eqnarray}
&\rho_\text{CW}(t+M)=\sum_{k=1}^M\Gamma_k U_\text{CW}^{(k)} \rho_\text{CW}(t+M-k) U_\text{CW}^{(k)\dag}.
\label{M2}
\end{eqnarray}
Again, $0\leqslant\Gamma_k\leqslant 1$ and $\sum_k\Gamma_k=1$. The
density matrix at each time $t>1$ will involve some mixing of the
different coins-walks, introducing correlations. If $M=1$ or
$\Gamma_1=1$, this reduces to the memoryless QRW. Like
Eq.~(\ref{M0}), the degree of correlations can be adjusted by
$\Gamma_k$. However, unlike the previous model, there is no
bifurcation or transition from bimodality to unimodality in the
behavior of the probability distribution vs $\Gamma$ --- see
Fig.~\ref{m2-combined}.


\textit{A simple harmonic oscillator model.}--- A more physically relevant picture
in which the roles of correlations and memory-dependence may be explained is through
the Hamiltonian formalism. To this end, keeping the discussion simple (for numerics), we modify the
``Hamiltonian" of the memoryless coined QRW model so that a history-keeping agent can
be added. This picture can easily be generalized to more realistic cases in which
the dynamics is continuous in time and also there is no need for an extra quantum
coin. The Hamiltonian of the memoryless QRW can be found from $U_{\text{CW}}(p)=
e^{i\frac{\pi}{2}}e^{iZ\otimes P}e^{-i\frac{\pi}{2}(\sqrt{1-p}X+\sqrt{p}Z)}$ by using
the Baker-Campbell-Hausdorff formula. Now consider that the walker is coupled to a simple
harmonic oscillator and a reservoir consisting of a sufficiently large number of excitation
modes in the following manner:
\begin{eqnarray}
& H_{\text{CWOR}}=H_{\text{CW}}+H_{\text{O}}+H_{\text{R}}+ H_{\text{OR}} +\lambda P(a+a^{\dag}).
\label{Hamiltonian}
\end{eqnarray}
Although this ``Hamiltonian" models a crude simplification, clearly
lacking various realistic characteristics, to some extent it has
been inspired by a cavity QED-based proposal for a QRW
~\cite{barry03}, and serves well enough for demonstration of our
ideas. In Eq.~(\ref{Hamiltonian}), $H_{\text{O}}=\frac{1}{2}\omega
a^{\dag}a$,
$H_{\text{R}}=\frac{1}{2}\sum_{k=1}^\infty\Omega_kc_k^{\dag}c_k$,
and for the coupling of the oscillator and reservoir degrees of
freedom we can take, for example,
$H_{\text{OR}}=g(a\sum_{k=1}^\infty
c_k^{\dag}+a^{\dag}\sum_{k=1}^\infty c_k)$, where $a$ is the
lowering operator of the oscillator
($a^{\dag}a|n\rangle_{\text{O}}=n|n\rangle_{\text{O}}$, for
$0\leqslant n\leqslant n_{\max}<\infty$) and $c_k$ is the
annihilation operator of the $k$th excitation mode. The
walker-oscillator coupling term, $H_{\text{WO}}=\lambda
P(a+a^{\dag})$, implies that as the walker moves over the lattice,
energy is being exchanged with the oscillator, and the
oscillator gets partial information about the walker. This
interaction is effectively a momentum-position coupling, which
preserves the walker's momentum. The history-dependence in this
model is modulated by the coupling constant $\lambda$ ($\lambda=0$
corresponds to the memoryless QRW). The coefficient $g$, instead,
controls the effect of the reservoir. The mediated coupling of the
oscillator to the reservoir may also result in long-term correlations
--- this, however, needs a closer analysis which is beyond the scope
of this paper.  Considering that
$[H_{\text{CW}},H_{\text{WO}}+H_{\text{O}}]=0$, we have
$U_{\text{CWO}}=U_{\text{CW}}e^{-i(H_{\text{WO}}+H_{\text{O}})}$.
To maintain tractability of the numerical simulations, here we make
the following assumptions: (i) ignoring the reservoir effect, i.e.,
$g=0$, (ii) taking a balanced coin-flip ($p=1/2$) and an initially
balanced coin, (iii) the walker is initially localized at the
center, (iv) $\omega=5$, (v) $T=60$, (vi) periodic boundary
condition, $S|L\rangle_{\text{W}}=|-L\rangle_{\text{W}}$, with
$L=75$, (vii) the oscillator is initially prepared at the ground state
$|0\rangle_{\text{O}}$, (viii) the oscillator energy levels (or
Hilbert space) are truncated at $n_{\max}=10$, and (ix) working in
$0\leqslant \lambda\leqslant 1$ interval for the coupling constant.
Validity of assumption (viii) is confirmed through the simulations
noting that after $T=60$ the maximum probability for
$|10\rangle_{\text{O}}$ being populated is of the order of
$10^{-24}$. Moreover, we have seen that taking $L=75$ and $T=60$
makes the boundary effects insignificant.
\begin{figure}[bp]
\includegraphics[width=8.5cm]{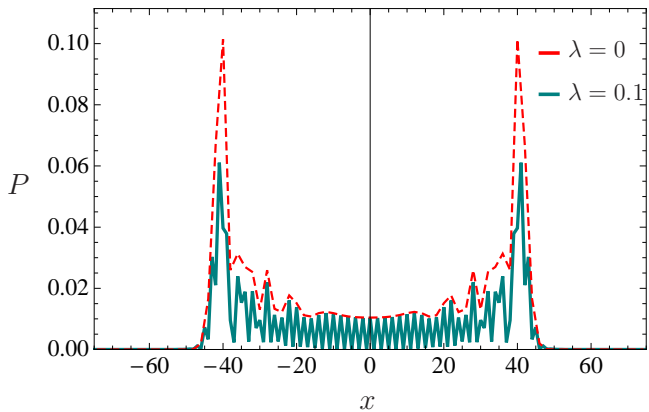}
\caption{(color online). Probability distribution for the
history-dependent QRW. $\lambda=0$ walk is only plotted at even
points and for $|x|\leqslant 60$, as it has zero probability at odd
points and for $|x|>60$. The  $\lambda=0.1$ walk has non-zero
probabilities everywhere on the lattice.} \label{probdist}
\end{figure}

\begin{figure}[bp]
 \includegraphics[width=8.5cm]{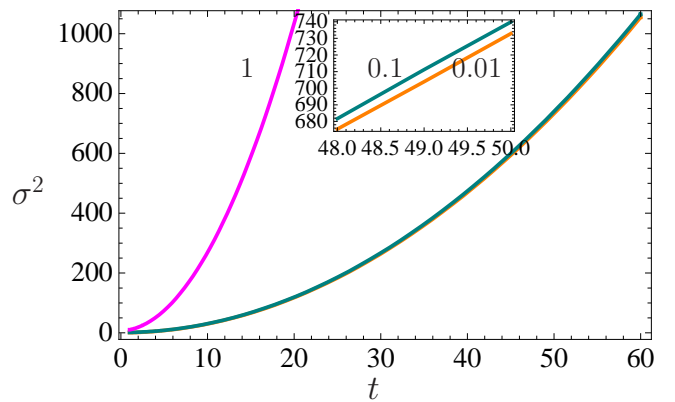}
\caption{(color online). Variance vs time for the modified QRW. The
inset highlights that $\sigma^2(0.1)>\sigma^2(0.01)$.} \label{variance}
\end{figure}

Figure~\ref{probdist} depicts the probability distribution for the
modified QRW model at $T=60$, in which symmetry is accounted for by
the symmetry of the initial conditions. The distribution with
$\lambda=0.1$ maintains a relatively similar behavior as the
memoryless QRW --- indicating a weak coupling regime --- whilst
there are non-zero probabilities everywhere on the lattice due to
the $\lambda$-coupling, indicating spreading of the distribution as
compared to the memoryless QRW. This can can be seen in
Fig.~\ref{variance}, variance vs time for three different values of
$\lambda$, in which a fitting of the form
$a(\lambda)t^2+b(\lambda)t+c(\lambda)$ yields $a(1)\approx 2.519$,
$a(0.1)\approx 0.295$, and $a(0)\approx 0.292$ --- a faster
spreading for the modified model, which may be understood by noting
that the $\lambda$-coupling favors hopping of the walker.

\begin{figure}[tp]
 \includegraphics[width=8.5cm]{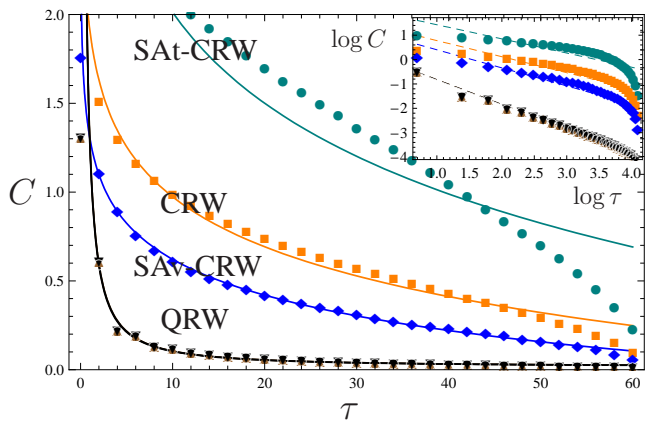}
\caption{(color online).  Autocorrelation function $C(\tau)$, for
modified QRW (for $\lambda=0,0.01,0.1$), self-avoiding
memory-dependent CRW (SAv-CRW), memoryless CRW, and self-attracting
memory-dependent CRW (SAt-CRW). Closer examination reveals that,
from top to bottom, the QRW plots are ordered $\lambda=0$,
$\lambda=0.01$, and $\lambda=0.1$.} \label{corr}
\end{figure}

\textit{Comparison of the walks.}--- To make a meaningful comparative analysis of how memory
and correlations behave in different RWs, we choose a memory-dependent CRW model proposed
in Ref.~\cite{tan02} (for other memory-dependent CRW models see, for example,
\cite{hara79,stanley83,hilfer91,hod05}). This model is interesting in that it features
various aspects of a real-world history-keeping systems, such as ``saturation'' of memory
(i.e., finite memory capacity). Unfortunately a direct quantum mechanical extension of
the model entails non-linearity and necessity of feedback. In this model, memory is based
on the ``information'' of the sites, which is determined by the number of the times the
walker visits each site as well as the times in which these visits occur. The walker
hops from site $i$ to a neighboring site $j$ with the probability $p_{ij}\sim
e^{u(s_j-s_i)}$, where $s_i(t)=\sum_mn_i(m)e^{-\kappa(t-m)}$ is the remaining
information at site $i$ at time $t$. Here $u$ is the density of information energy;
$u>0$ ($u<0$) corresponds to a walker attracted to (repelled by) sites with high
information content and $u=0$ gives the memoryless CRW. The coefficient $\kappa\geqslant
0$ is the memory decay exponent, for our simulations fixed at $\kappa=10^{-4}$, and $n_i(m)$
is $1$ if site $i$ was visited at time $m$ and $0$ otherwise. A saturated information
amount, $s_{\max}$, is assumed above which the effect of information ceases to increase
($s(t)\leqslant s_{\max}$), which for our simulations is taken to be $13$. The effect
of memory on the scaling exponents of the RW, in terms of the number of sites visited
and the distance the walker travels from its initial position, has been examined
\cite{tan02}, demonstrating that for any $\kappa>0$ and finite $u$ this RW exhibits a
similar scaling behavior of variance to that of a memoryless CRW. As $u\rightarrow\infty$,
the scaling behavior of the variance of this model changes from $\sigma^2\sim T$ to
$\sigma^2\sim c$, where $c$ is a constant; whilst for $u\rightarrow-\infty$, the scaling
changes from $\sigma^2\sim T$ to $\sigma^2\sim T^2$. For our numerical analysis $u$ remains
restricted such that the variance behaves as expected for CRW.
The probability distribution for this memory-dependent CRW is calculated using $10^4$ independent
repetitions and an averaging over the results.

\begin{figure}[tp]
 \includegraphics[width=8.5cm]{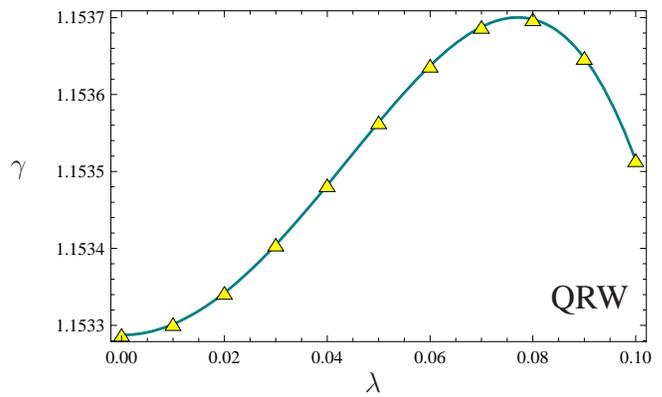}
\caption{(color online). Plot of $\gamma$ vs $\lambda$ for the
history-dependent QRW. $\gamma$ increases with $\lambda$, as the
peak and subsequent decrease is a consequence of the finite size of
our walker's space.} \label{gplot}
\end{figure}

A standard method to characterize short- and/or long-term
memory-dependent behaviors in data analysis is through calculation
of (auto-) correlation functions \cite{honerkamp}. In the case of
quantum diffusion systems, it has also been argued that the temporal
scaling behavior of the correlation function can show some universal
characteristic relation with the spreading and the spectrum
\cite{decay,zhong95,spreading,zhong-prl-01,mukho07,ribeiro04}. We
adopt a modified definition for the correlation function as follows.
From the probabilities at a fixed $x^*$ for varying time $t$ (up to
$T\lesssim \infty$), the time-series $\lbrace P_{x^*}(t)\rbrace_{t=0}^T$
is generated. Next, we define $C_{x^*}(\tau)\equiv\sum_{t=0}^{T-\tau}
P_{x^*}(t)P_{x^*}(t+\tau)$. An advantage of this definition, to the one used in
Refs.~\cite{decay,zhong95,spreading,zhong-prl-01}, is that ours
allows to build time-series data for a QRW in the same footing as
CRW. Besides, it enables application of (classical) data analysis
tools, e.g., detrended fluctuation analysis \cite{hu01}, for finding
trends and fractal behaviors in a data set. It has been shown that
the correlation function exhibits a power-law (i.e., algebraic) decay as
$C(\tau)\sim\tau^{-\gamma}$, where $\gamma$ --- the correlation
exponent --- is related to spectral properties of the system
\cite{zhong95}. A small value of $\gamma$ is an attribute of a walk
that stays relatively localized, whilst a large $\gamma$ indicates a
tendency for the walker's distribution to spread with time. The
correlation function has been calculated for the memory-dependent
CRW (for $u=\lbrace0,\pm 0.1\rbrace$) and the modified QRW (for
$\lambda=\lbrace0,0.01,0.1\rbrace$) at $x^*=0$. The value of
$C(\tau)$ is very small for odd values of $\tau$. This is because
the probability of the walker occupying the origin in odd $t$s is
small. These probabilities are non-zero in the case of $\lambda>0$
as seen earlier. In CRW and for $\lambda=0$, this probability is
zero. The even (time) points of the correlation function have been fitted
to $C(\tau)\sim a+b\tau^{-\gamma}$ --- Fig.~\ref{corr} --- with
$\gamma_{\text{QRW}}\approx 1.153$, $\gamma_{\text{SAv-CRW}}\approx
0.01$, $\gamma_{\text{CRW}}\approx 0.005$, and
$\gamma_{\text{SAt-CRW}}\approx 0.003$. The behavior of $\gamma$
vs $\lambda$ has been plotted in Fig.~\ref{gplot}. A preliminary analysis suggests that
the peak in this plot may be a byproduct of the finite size of the system. A closer
investigation, however, may suggest a better explanation for possible
underlying reason(s). The property $\gamma_{u=0.1}<\gamma_{u=0}
<\gamma_{u=-0.1}$ is in accordance with our understanding about the physical
meanings of positive and negative $u$s, i.e., self-attracting and self-avoiding,
respectively. An interesting observation is that in the limit of
$u\rightarrow -\infty$, SAv-CRW and QRW show similar general
behaviors.

\textit{Summary.}--- We have introduced simple quantum random walk
models with memory-dependent features by adding a non-Markovian
property to the walk as an uncorrelated mixing of the states at
different instants and also by introducing a Hamiltonian picture
with a memory modulating coupling. Variance vs time for the
uncorrelated models has been calculated. In the Hamiltonian model,
we have defined and calculated the concepts of correlation and
correlation exponent as useful tools for assessing the effect of
memory or correlation. Comparison with classical memoryless and memory-dependent
models has indicated an anti-correlation in the quantum random walk.
Variance as an indicator to distinguish between classical and
quantum regimes has apperared to be a not so useful tool.
We, instead, have suggested that tools such as correlation exponents
and detrended fluctuation analysis are probably more useful in characterizing different regimes of a
quantum system. These studies may shed some light on how different
regimes of behaviors in quantum diffusion systems emerge and how
they are related to other physical characteristics of those systems.
Also, these might have some implications on when a quantum random
walk based algorithm for a problem may result in a speedup in
comparison to classical algorithms.

\textit{Acknowledgements.}--- This work was supported by CIFAR, \textit{i}CORE, MITACS, NSERC, and PIMS.


\end{document}